# Mn Interstitial Diffusion in GaMnAs

K.W. Edmonds[1], P. Bogusławski[2,3], B.L. Gallagher[1], R.P. Campion[1], K.Y. Wang[1], N.R.S. Farley[1], C.T. Foxon[1], M. Sawicki[2], T. Dietl[2], M. Buongiorno Nardelli[3], J. Bernholc[3]

[1]School of Physics and Astronomy, University of Nottingham, Nottingham NG7 2RD, United Kingdom
[2]Instytut Fiziki PAN, PL-02668 Warzawa, Poland
[3]North Carolina State University, Raleigh, USA

*We present a combined theoretical and experimental study of the ferromagnetic semiconductor GaMnAs which explains the remarkably large changes observed on low temperature annealing. Careful control of the annealing conditions allows us to obtain samples with ferromagnetic transition temperatures up to 160 K. Ab initio calculations, and measurements of resistivity changes during annealing, show that the observed changes are due to out-diffusion of Mn interstitials toward the surface, governed by an energy barrier of about 0.7-0.8 eV. The Mn interstitial reduces the hole density and couples antiferromagnetically with substitutional Mn, thereby suppressing the ferromagnetism. Electric fields induced by high concentrations of substitutional Mn acceptors have a significant effect on the diffusion barriers.*

The III-V dilute magnetic semiconductors (DMS), in which high levels of transition metal impurities are incorporated into a semiconductor host, offer good prospects for the effective integration of ferromagnetic properties into semiconductor heterostructures [1]. Here, a long range ordering of the magnetic dopants is mediated by the charge carriers, leading to the possibility of spin engineering by manipulation of the free carriers [2,3]. Interest in these materials was further stimulated by predictions of ferromagnetism at room temperature and above [4]. Many of the physical properties of DMS materials have been successfully explained within the Zener model of carrier-mediated ferromagnetism [4-6], which predicts a ferromagnetic transition temperature $T_C$ determined by the concentration of holes $p$, and roughly proportional to $p^{1/3}$.

The (Ga,Mn)As system provides a valuable test ground for the properties of DMSs, due to the relatively high $T_C$ and integration with the well-characterised GaAs system. The Mn dopant substitutes for the Ga site, and fulfils two roles: it provides a local spin 5/2 magnetic moment, and acts as an acceptor, providing itinerant holes which mediate the long range magnetic order. Practical applications of this material will require an increased control of $T_C$. An important breakthrough was the discovery that annealing at temperatures comparable to [7-10] or lower than [11-12] the growth temperature can result in dramatic enhancements of $T_C$ and $p$, as well as the saturation magnetisation. GaMnAs is grown at relatively low substrate temperatures (~200-250ºC) in order to achieve above equilibrium concentrations of Mn in the GaAs host. This tends to lead to high defect concentrations, the most important of which are As-antisite defects, $As_{Ga}$, and Mn interstitials, $Mn_I$, both double donors that compensate the carriers provided by substitutional $Mn_{Ga}$. The increase of $T_C$ may therefore be related to a removal of these defects. Indeed, ion channelling experiments have indicated a decreased concentration of $Mn_I$ in the annealed material [10]. However the underlying mechanism that leads to such large changes at temperatures as low as 180ºC [12] has not previously been determined.

Here we report on a combined experimental and theoretical study of the microscopic annealing processes in GaMnAs. *In-situ* monitored resistivity measurements as a function of film thickness are shown to be consistent with out-diffusion of compensating defects. Using *ab initio* analysis we elucidate the possible pathways for $Mn_I$ interstitial diffusion, considering the effect of $Mn_{Ga}$-$Mn_I$ complexes and electric fields induced by the high concentrations of $Mn_{Ga}$ acceptors. The calculated energy barrier is in good agreement with our measured value of $(0.7\pm0.1)$eV. These results offer new insights into the role of defects in GaMnAs and the optimisation of material properties.

Growth of $Ga_{1-x}Mn_xAs$ films is described in detail in ref. [13]. The films are grown on GaAs(001) substrates by low temperature (≈200ºC) molecular beam epitaxy. The Mn concentration was $x=0.067$ [14], which was found in our previous study to give the highest values of $T_C$ [12]. Post-growth annealing was performed while simultaneously measuring the electrical resistance in a van der Pauw geometry. $T_C$ was obtained from extraordinary Hall effect measurements [11,12] as well as from the temperature dependence of the remnant magnetisation measured by SQUID magnetometry. The two methods give the same values for $T_C$ within ±2K.

The *in-situ* monitored resistivity during annealing at 190ºC is shown in figure 1a, for films of thickness 10nm, 25nm, 50nm, 100nm. The as-grown samples at time=0 have the same resistivity within 10%. The resistivity then falls as the samples are annealed. A decrease of the resistivity of (Ga,Mn)As films is typically observed on low-temperature annealing [7-12]. This is predominantly due to an increase of the carrier density (the carrier density increases by ≈100% for $x=0.067$, while mobility changes are <≈10% [11]). The resistivity falls more rapidly with decreasing film thickness, indicating that the ability to deactivate compensating defects is a strong function of the thickness. This suggests that diffusion of $Mn_I$ to the surface or into the substrate is the dominant mechanism in removing the defects, rather than the formation of random precipitates postulated in ref. [10].

The data of Fig. 1a can be interpreted in terms of one-dimensional out-diffusion of mobile compensating defects from a layer of thickness $L$. The defect density at depth $x$, time $t$ is given by [15]:

$$n'(x,t) = \frac{N}{L}(4\pi Dt)^{-0.5} \int_{-L/2}^{L/2} \exp\left(\frac{-(x-x')^2}{4Dt}\right) dx'$$

where $N$ is the total number of defects per unit area and $D$ the diffusion coefficient. The number of defects per unit area within the layer at time $t$ is then given by:

$$n(t) = \int_{-L/2}^{L/2} n'(x,t) dx$$

If the change in resistivity is entirely due to the increase in carrier concentration caused by the removal of compensating $Mn_I$, the time-dependent resistivity can be modelled using:

$$\rho(t) = (\sigma_0 - \sigma_1 n(t))^{-1}$$

with $\sigma_0$, $\sigma_1$ and $D$ as fit parameters. The model reproduces the experimental data shown in Fig. 1a. In Fig. 1b the rate of change of conductivity, $d(1/\rho(t))/dt$, is plotted versus $t/L^2$. The data for different thicknesses fall on a single universal curve, which is in excellent agreement with this 1D diffusion picture.

In practice, diffusion into the substrate is likely to be limited by electrostatics, as a *p-n* junction will eventually form between ionised $Mn_I$ donors and $Mn_{Ga}$ acceptors. In contrast, diffusion to the surface will lead to passivation by e.g. oxidation, so this is the most likely mechanism for removal of $Mn_I$. This may explain why low temperature annealing is inefficient in (Ga,Mn)As films with capping layers [16]. Consistent with this, we observe an increased concentration of surface Mn in annealed samples by Auger spectroscopy. The detailed diffusion mechanism will affect the distribution of $Mn_I$ during annealing, but not the resistivity drop and its $L^2$-dependence.

The highest $T_C$ obtained after annealing is shown in the inset of Fig. 1b. $T_C$ shows relatively weak thickness dependence, with a maximum value of 159K for the 25nm thick film. In Ref. [8], it was reported that $T_C$ is limited to 110K for films thicker than ~60nm, a value obtained by other groups [7-10] and suggested to be a fundamental limit [10]. In the present study, $T_C$ is significantly higher than 110K even for the 100nm thick film. The annealing in ref. [8] was performed at higher temperatures (≈250ºC), and annealing for longer than ~3 hours led to a reduction of $T_C$ [7]. The mechanism leading to a decreased $T_C$ is presently unknown, but the lower temperature employed here seems to mostly inhibit this. It is clear from figure 1a that the removal of compensating defects from the GaMnAs layer requires long anneal times at 190ºC, and also that the resistivity of the 100nm film is still slowly falling even after 250 hours. Therefore, the thickness dependence of $T_C$ observed in the present study and elsewhere may simply be due to an incomplete out-diffusion in the thicker layers.

Annealing at different temperatures allows us to determine the temperature dependence of the diffusion coefficient, and thus the energy barrier $Q$ governing the diffusion process, from the relationship $D=D_0 exp(-Q/kT)$. 25nm thick (Ga,Mn)As films from the same wafer were annealed at temperature T between 160°C and 200°C, while monitoring the resistivity *in-situ* as above. The value of $D$ obtained from the above fitting procedure is plotted versus 1/T in Fig. 2. From this we obtain $Q$=(0.7±0.1) eV.

We now compare these experimental results with *ab initio* calculations. These were performed within the Local Spin Density Approximation, using the plane wave code developed in Trieste [17]. We used a large unit cell with 64 atoms in the ideal case. Thus, substitution of one Ga atom by Mn corresponds to the alloy containing 3.1 % of Mn, etc. The Brillouin zone summations, performed using the Monkhorst-Pack scheme with 4 points in the irreducible part of the ideal *folded* Brillouin Zone, gave convergent results. Positions of all atoms in the unit cell were allowed to relax, which is particularly important for a correct evaluation of diffusion barriers. In actual samples all $Mn_I$ double donors are ionized to the charge state 2+, since the concentration of $Mn_I$ is significantly lower than that of $Mn_{Ga}$ acceptors. Accordingly, in the calculations the Fermi level is fixed to reflect this situation.

In the zinc blende structure there are two interstitial sites with tetrahedral coordination, the first of which, T:Ga, is surrounded by 4 Ga, and the second, T:As, by 4 As atoms. Both sites are shown in figure 3. In the ideal case, the distance from the T site to a nearest neighbour (NN) atom is equal to the bond length of the host crystal. The electronic structure of $Mn_I$ at both sites is very similar, and the results for T:As agree well with those calculated elsewhere [18]. In both cases, the majority d(Mn) spin-up states form a narrow band located about 0.3 eV above the top of the valence band and occupied with 5 electrons, while the minority d(Mn) spin-down states form a band in the vicinity of the bottom of the conduction band. Thus, in both locations $Mn_I$ is a double donor. Interstitial diffusion proceeds along …-T:Ga-T:As-T:Ga-… paths, where individual jumps may occur both in the (110) and in the orthogonal (1$\bar{1}$0) plane (in a 'ball-and-stick' model, the network of interstitial sites is an exact replica of the zinc blende lattice). By symmetry, it is sufficient to investigate one such segment. The calculated changes of the total energy of $Mn_I$ as a function of its position along the T:Ga-T:As path are presented in Fig. 4. The equilibrium position of an isolated $Mn_I^{2+}$ is the T:As site, where its energy is lower than at T:Ga by 0.35 eV. The stabilization of $Mn_I$ at the T:As site is due to the fact that the positively charged $Mn_I^{2+}$ is attracted by negatively charged As anions, and repelled by positively charged Ga cations. We stress here, that not only this result, but also other aspects of diffusion discussed in the following, are largely determined by Coulomb interactions of $Mn_I$ with acceptors, anions, and cations. The calculated energy barrier for diffusion of $Mn_I^{2+}$ is 0.8 eV, in good agreement with the measured value.

Since there exists a Coulomb attraction between ionized donors and acceptors, we also consider formation of $Mn_I$-$Mn_{Ga}$ pairs and $Mn_{Ga}$-$Mn_I$-$Mn_{Ga}$ complexes. They are coupled by both Coulomb and magnetic interactions. The formation of stable pairs may in principle render the annealing less efficient and block the out-diffusion of $Mn_I$. We first examine the binding of these complexes. Let us first assume that the $Mn_{Ga}$ occupies the site 'a' in Fig. 3. We find that there are two configurations of the $Mn_I$-$Mn_{Ga}$ NN pair which have the same energy, with $Mn_I$ occupying the T:Ga or T:As site, respectively. The T:As is more distant from $Mn_{Ga}$ than T:Ga, but the decreased attraction with $Mn_{Ga}$ is compensated by the increased attraction to the nearest As neighbors. There are 4 T:Ga and 6 T:As NN sites. The energy barrier between the two sites is small, 0.5 eV, so that $Mn_I$ will frequently swap between them even at room temperature. The calculated total energy change as a function of the distance of $Mn_I$ from $Mn_{Ga}$ is shown in figure 4. The considered distance corresponds to the jump of $Mn_I$ from T:Ga to the T:As2 site in Fig. 3. An antiferromagnetic orientation of Mn magnetic moments was assumed (see below). It follows from figure 4 that the $Mn_I^{2+}$ at T:Ga indeed forms a NN pair with $Mn_{Ga}$, mainly because of the Coulomb attraction between $Mn_I^{2+}$ and $Mn_{Ga}^-$. The calculated energy barrier for the jump of $Mn_I^{2+}$ from the T:Ga to the T:As2 site is 1.3 eV. This jump may be regarded as activation of the dissociation of the NN pair. A second possible dissociation path begins with a T:As-T:Ga2 jump, see Fig. 3; the corresponding energy barrier is 1.1 eV, i.e., lower than in the former case.

The geometry analyzed above corresponds to a good approximation to the case of an isolated $Mn_{Ga}$, because of relatively large distances between $Mn_{Ga}$ ions in the 64-atom unit cell. On the other hand, investigated samples contain 6.7 % of Mn, and in this case there are (in average) 2 $Mn_{Ga}$ ions per 64 atoms. As we will now show, electric fields induced by this high density of acceptors lower the diffusion barriers.

We first assume that ionized $Mn_{Ga}$ ions avoid each other. Accordingly, we place $Mn_{Ga}$ as far apart as possible in our 64-atom cubic cell (sites 'a' and 'd' in Fig. 3). Due to the presence of the second acceptor at 'd', the calculated barrier along the path T:Ga-T:As2 is reduced from 1.3 to 1.0 eV. Moreover, the distances of $Mn_I$ at T:As2 to the 'a' and 'd' sites are equal, which implies an equal Coulomb attraction by both acceptors. If the second $Mn_{Ga}$ acceptor is located somewhat closer (e.g., at site 'c'), the dissociation barrier along the T:Ga-T:As2 path is reduced even more. Clearly, a similar decrease of the barrier by about 0.2 eV is expected for the second path, T:As-T:Ga2.

In the opposite limit, two $Mn_{Ga}$ atoms are located close to each other, i.e., at sites 'a' and 'b' respectively. In a stable complex, a $Mn_{Ga}$-$Mn_I$-$Mn_{Ga}$ triangle is formed, with $Mn_I$ at T:Ga. In the first step of dissociation, $Mn_I$ jumps from T:Ga to T:As2 with the barrier of 1.25 eV. The subsequent jumps to more distant sites require overcoming barriers of about 1.1 eV.

In actual samples, $Mn_{Ga}$ ions are distributed randomly, and diffusion has a character of percolation process with a spread of energy barriers. A more detailed treatment of this complex problem is beyond the scope of this paper. However, it is clear that the electric fields induced by a high density of acceptors lower the activation barrier for diffusion from 1.1 eV to about 0.9±0.1 eV, leading to satisfactory agreement with the measured value of 0.7±0.1 eV. Moreover, one should expect that in samples with a lower Mn content, e.g. 1%, the barriers are higher, and the annealing is less efficient.

To evaluate the magnetic coupling of the $Mn_{Ga}$-$Mn_I$ pair we compare the energies of the ferromagnetic (FM) and antiferromagnetic (AF) orientations of their magnetic moments. For both T:Ga and T:As locations of $Mn_I$, the energy difference between the two orientations, $\Delta E^{A-F}$, is -0.5 eV, which corresponds to an antiferromagnetic interaction. This AF character has been anticipated in Ref. [19] based on model tight binding calculations. The coupling has a short-range character, since $\Delta E^{A-F}$ vanishes for the second NN (i.e., for T:As2 or T:Ga2 sites) and more distant pairs. This behavior is in sharp contrast with the long-range interaction between a pair of substitutional Mn ions. Turning to the $Mn_{Ga}$-$Mn_I$-$Mn_{Ga}$ triangle, in the magnetic ground state the moments of the two substitutional $Mn_{Ga}$ are parallel since the $Mn_{Ga}$-$Mn_{Ga}$ pair is coupled ferromagnetically, while the orientation of the $Mn_I$ moment is antiparallel, in agreement with the AF $Mn_{Ga}$-$Mn_I$ coupling. The FM orientation of the three magnetic moments of the complex is 0.65 eV higher in energy than the ground state configuration. An AF coupling of a significant fraction of interstitial and substitutional Mn may explain why the measured magnetic moment of (Ga,Mn)As is often smaller than the predicted value of $4\mu_B$ per Mn [20], and why this increases on removal of $Mn_I$ by annealing [9].

Finally, and importantly, the kick-out mechanism of diffusion ($Mn_I$+$Ga_{Ga}$ --> $Mn_{Ga}$+$Ga_I$) is non-efficient, since the calculated barrier is about 3 eV.

In summary, a combined experimental and theoretical study of (Ga,Mn)As thin films has demonstrated that the strong increases of the Curie temperature (up to 159 K), carrier density and saturation magnetisation result from the out-diffusion of interstitial Mn ions towards the surface. The diffusion is affected by electric fields of ionized Mn acceptors. The effective diffusion barrier is about 0.7-0.8 eV.

We thank Z. Wilamowski and J.L. Dunn for enlightening discussions. This work is supported by FENIKS project (EC:G5RD-CT-2001-00535), Ohno Semiconductor Spintronics ERATO project of JST, grant PBZ-KBN-044/P03/2001 and grants from US ONR and DoE.

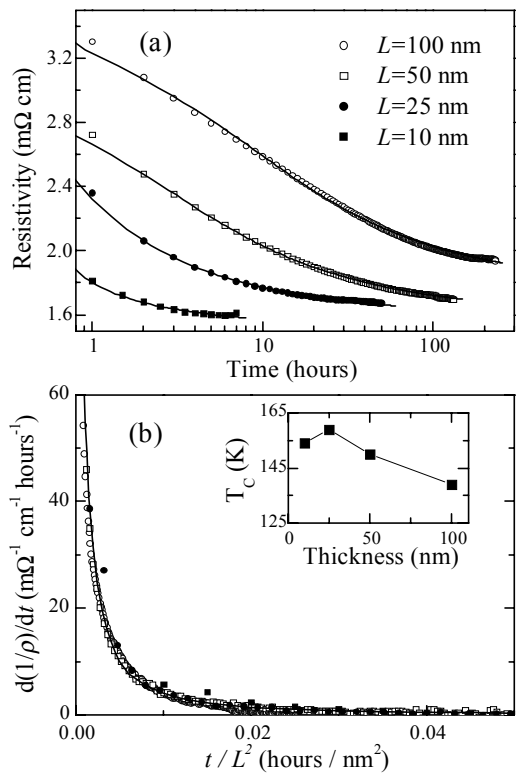

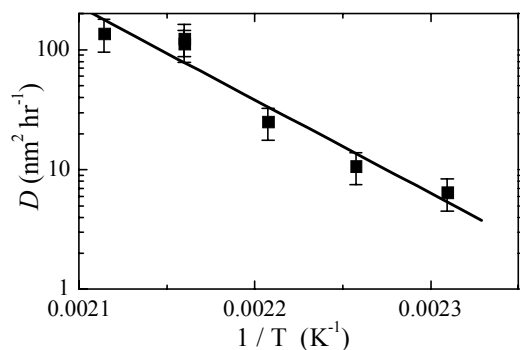

Figure 1. (a) *In-situ* monitored resistivity versus annealing time $t$, for $Ga_{0.933}Mn_{0.067}As$ films of thickness $L$=100nm (open circles), 50nm (open squares), 25nm (closed circles), and 10nm (closed squares), as well as fits using a 1D diffusion model; (b) rate of change of conductivity versus $t/L^2$ for the data in (a); inset of (b): ferromagnetic transition temperature of annealed films versus thickness.

Figure 2. Diffusion coefficient for 25nm thick $Ga_{0.933}Mn_{0.067}As$ films versus $1/T$, for temperature T between 160°C and 200°C.

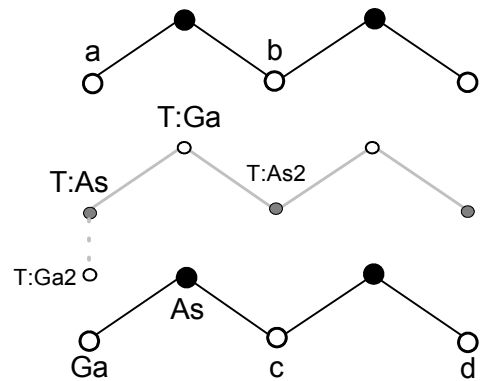

Figure 3. Projection of atomic positions on the (110) plane.

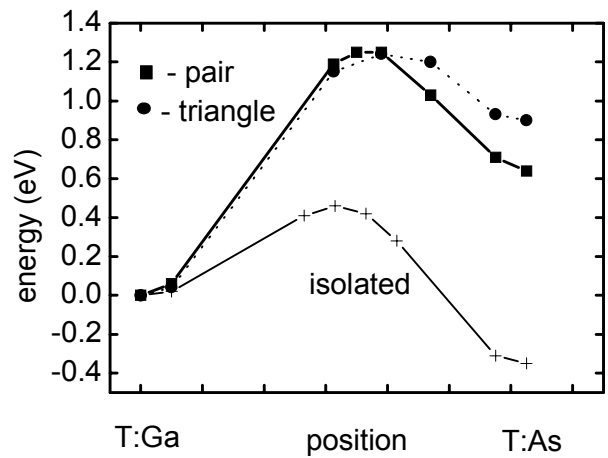

Figure 4. Calculated changes of the total energy of $Mn_I$ along the T:Ga-T:As path for isolated $Mn_I$, $Mn_I$-$Mn_{Ga}$ pair, and $Mn_{Ga}$-$Mn_I$-$Mn_{Ga}$ triangle.